%% file: main.tex
\documentclass[conference]{IEEEtran}
\IEEEoverridecommandlockouts
\usepackage{cite}
\usepackage{amsmath,amssymb,amsfonts}
\usepackage{bm}
\usepackage{graphicx}
\usepackage{textcomp}
\usepackage{xcolor}
\def\BibTeX{{\rm B\kern-.05em{\sc i\kern-.025em b}\kern-.08em
    T\kern-.1667em\lower.7ex\hbox{E}\kern-.125emX}}

\input{custom-style.tex}

\usepackage{tikz}
\usepackage{subfig}
\usepackage{array}
\newcolumntype{P}[1]{>{\centering\arraybackslash}p{#1}}
\begin{document}

\title{ Efficient Quantum Counting and Quantum Content-Addressable Memory for DNA similarity }

\author{
\IEEEauthorblockN{Jan Balewski\textsuperscript{*}}
\IEEEauthorblockA{\textit{National Energy Research Scientific Computing Center} \\
\textit{Lawrence Berkeley National Laboratory}\\
Berkeley, CA, USA  \\
balewski@lbl.gov}
\and
\IEEEauthorblockN{Daan Camps\textsuperscript{*}}
\IEEEauthorblockA{\textit{National Energy Research Scientific Computing Center} \\
\textit{Lawrence Berkeley National Laboratory}\\
Berkeley, CA, USA  \\
dcamps@lbl.gov}
\and
\IEEEauthorblockN{Katherine Klymko}
\IEEEauthorblockA{\textit{National Energy Research Scientific Computing Center} \\
\textit{Lawrence Berkeley National Laboratory}\\
Berkeley, CA, USA  \\
kklymko@lbl.gov}
\and
\IEEEauthorblockN{Andrew Tritt}
\IEEEauthorblockA{\textit{Applied Mathematics and Computational Research Division} \\
\textit{Lawrence Berkeley National Laboratory}\\
Berkeley, CA, USA \\
ajtritt@lbl.gov}

\and
\\
{\footnotesize \textsuperscript{*}equally contributing authors}
}
\maketitle

\begin{abstract}

We present QCAM, a quantum analogue of Content-Addressable Memory (CAM),
useful for finding matches in two sequences of bit-strings.
Our QCAM implementation takes advantage of Grover's search algorithm
and proposes a highly-optimized quantum circuit implementation of the
QCAM oracle.
Our circuit construction uses the parallel uniformly controlled rotation gates, which were used in previous work to generate QBArt encodings.
These circuits have a high degree of quantum parallelism which reduces their critical depth.
 The optimal number of repetitions of the Grover iterator used in QCAM depends on the number of true matches and hence is input dependent.
We additionally propose a hardware-efficient implementation of the quantum counting algorithm (HEQC) that can infer the optimal number of Grover iterations from the measurement of a single observable.
We demonstrate the QCAM application for computing the Jaccard similarity between two sets of k-mers obtained from two DNA sequences.

\end{abstract}




\section{Introduction}

The classical content-addressable memory (CAM)~\cite{cam}, also called  associative memory,  is a special purpose memory circuit that implements a lookup table function in a single clock cycle. CAM compares input search data against a table of stored data and returns the address of matching data.
CAM has found applications in network routing as well as other areas.

In this paper, we design and evaluate a highly-optimized quantum circuit leveraging  the Grover oracle~\cite{grover} that implements the CAM lookup algorithm on quantum hardware and achieves a quadratic speedup over brute-force black box search. We call this application of Grover's algorithm \emph{QCAM}.
Compared to classical CAM relying on additional, specialized circuitry, QCAM runs on a standard gate-based quantum device. 
Furthermore, we take advantage of the parallel uniformly controlled rotation ({\it pUCR}) encoding algorithm~\cite{qbart} to harness the exponential dimensionality of the  Hilbert space and represent exponentially sized data sets on a polynomial number of qubits 
with a modest constant prefactor and a high degree of quantum parallelism.

A key hyperparameter  that is required for every Grover problem, including QCAM, is  knowledge of the optimal number of repetitions of the Grover iterator such that a solution is found with probability $O(1)$. The number of iterations to take depends on the size of the search space, which is typically known, and on the number of solutions to the Grover problem. The latter is typically not known in advance but can be inferred through a separate  quantum counting algorithm~\cite{qcounting}.
In our current work, we present a novel, hardware efficient implementation of such a quantum counting algorithm which bypasses the need for full quantum phase estimation and determines the number of iterations for the QCAM problem from measuring a single observable of a different circuit containing just one Grover oracle. 

We illustrate the use of QCAM for finding the intersection of two DNA sequences, each
represented as a sequence of overlapping $k$-mers.
This is the computationally most expensive step in calculating the {\it Jaccard similarity metric}~\cite{Jaccard} between two DNA sequences.
Previously proposed quantum pattern matching algorithms are also based on Grover search~\cite{QuRAM,Niroula}. Compared to these approaches, our work achieves a highly-optimized
quantum circuit implementation that has already been experimentally demonstrated on quantum hardware~\cite{qbart}. In combination with the optimized quantum counting algorithm, this opens a pathway to compute the Jaccard metric on current quantum hardware.
 
The remainder of this paper is organized as follows.
Sec.~\ref{sec:qbart} summarizes our previous results for the \qbart\ data encoding with pUCR gates~\cite{qbart}.
Sec.~\ref{sec:qcam} describes how we can use the pUCR circuits to construct
an end-to-end implementation of a Grover oracle that finds matches in two sequences of bit-strings.
Sec.~\ref{sec:HEQC} discusses the hardware efficient implementation of the quantum counting algorithm useful for inferring the optimal number of Grover iterations for QCAM.
Sec.~\ref{sec:dna} presents how to use QCAM for computing the Jaccard similarity metric between two DNA sequences.
We conclude in Sec.~\ref{sec:disc}.

\section{Encoding data sequences on a QPU}
\label{sec:qbart}


The \qbart~\cite{qbart} circuit is a highly-optimized circuit construction to prepare an NEQR~\cite{neqr} data encoding,
\begin{equation}
\ket{\psi(\mathbf{y})} = \frac{1}{\sqrt{N}} \sum_{i \in [N]} \ket{i} \otimes \ket{y_i},
\label{eq:neqr}
\end{equation}
for an ordered sequence of $N$ bit-strings, $\mathbf{y} = \left[y_0, \cdots, y_{N -1} \right]$, $N \equiv 2^{n}$,
where each bit-string $y_i$ consists of $d$ bits (bit-depth).
The original NEQR circuit~\cite{neqr} requires $d$ qubits for the data and $n$ qubits for the address and its  critical circuit depth is $2^{2n}d$. A much shallower  \qbart\ circuit leverages two circuit optimizations:
\begin{enumerate}
    \item[(1)] it uses uniformly controlled rotation (UCR) gates~\cite{ucr}, first used in the context of sequence encodings by QPIXL~\cite{qpixl} -- which reduces the depth to $2^{n} d$; and
    \item [(2)] permutations of the UCR gates are braided together in a \emph{parallel} UCR gate (pUCR) -- this further reduces the critical depth to $\lceil 2^{n} d / \min(n,d)\rceil$ of cycles with entangling gates.
\end{enumerate}

\begin{figure}[htbp!]
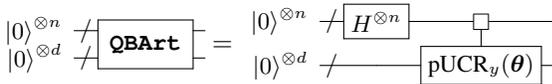

\centering
\(\small
\qquad \quad
\begin{myqcircuit}
\lstick{\ket{0}^{\otimes n}} & {/} \qw & \multigate{1}{\qbart} & \qw\\
\lstick{\ket{0}^{\otimes d}} & {/} \qw & \ghost{\qbart} & \qw\\
\end{myqcircuit}
= \qquad \quad \
\begin{myqcircuit}
\lstick{\ket{0}^{\otimes n}} & {/} \qw & \gate{H^{\otimes n}} & \ctrlsq{1} & \qw\\
\lstick{\ket{0}^{\otimes d}} & {/} \qw & \qw & \gate{\pUCRy(\bm{\theta})} & \qw\\
\end{myqcircuit}
\)
\caption{High-level block diagram of the \qbart\ circuit  encoding  a sequence of $2^{n}$ bit-strings of bit-depth $d$ onto the
$n+d$ qubit state \ket{\psi(\mathbf{y})} of Eq.~\eqref{eq:neqr}.}
\label{circ:hlqbart}
\end{figure}

This circuit depth reduction results in a high degree of quantum-parallel gate operations in pUCR gates.
Mathematically, a pUCR gate implements the unitary,
\begin{equation}
\begin{split}
\text{pUCR}_y(\bm{\theta}) \ & \ket{i} \ \ket{j_0} \otimes \cdots \otimes \ket{j_{d-1}}
\mapsto \\
& \ket{i} \ R_y(\theta_{i,0}) \ket{j_0} \otimes \cdots \otimes R_y(\theta_{i,d-1}) \ket{j_{d-1}},
\end{split}
\label{eq:pucrmat}
\end{equation}
with rotation angles $\bm{\theta} \in \mathbb{R}^{2^n \times d}$.
The high-level circuit diagram to prepare a \qbart\ NEQR encoding~\eqref{eq:neqr} using a pUCR gate is shown in Fig.~\ref{circ:hlqbart}.
In the \qbart\ circuit,
the Pauli-Y rotation angles, $\bm{\theta}$, in Eq.~\eqref{eq:pucrmat}
are chosen as
\begin{equation}
\theta_{i,j} =
\begin{cases}
0, & \text{if bit $y_{i,j} = 0$}, \\
\pi, & \text{if bit $y_{i,j} = 1$},
\end{cases}
\end{equation}
where $y_{i,j}$ is the $j$th bit of the $i$th bit-string in $\mathbf{y}$.

See~\cite{qbart} for details on the gate-level circuit implementation of the pUCR gate and its experimental realization on quantum hardware platforms including IBMQ, IonQ, and Quantinuum H1-1.

\section{QCAM: Constructing Grover search oracles from pUCR gates}
\label{sec:qcam}
In this section, we show how to leverage  pUCR gates to construct an efficient Grover search oracle $G$
for QCAM applications (data matching).
We assume the input data consists of two sequences of bit-strings,
\begin{equation}
    \begin{split}
        \mathbf{a} &= \left[a_0, \cdots, a_{N-1} \right], \\
        \mathbf{b} &= \left[b_0, \cdots, b_{M-1} \right],
    \end{split}
\end{equation}
that can be of different lengths $N$ and $M$, but both contain bit-strings of the same
bit-depth $d$.

In QCAM, we want to find either \emph{all} or \emph{a subset of} the pairs of addresses
$(i, j)$, $i \in [N]$, $j \in [M]$, for which $a_i = b_j$.
We assume throughout the remainder of the paper that both $N$ and $M$
are powers of 2 and write $N \equiv 2^n$, $M \equiv 2^m$. If the input
data sets don't satisfy this constraint, we can simply pad them to the next power of 2
in such a way that no additional, spurious matches are introduced. If required, this can
always be achieved by increasing the bit-depth to $d+1$.

\begin{figure}[htbp]
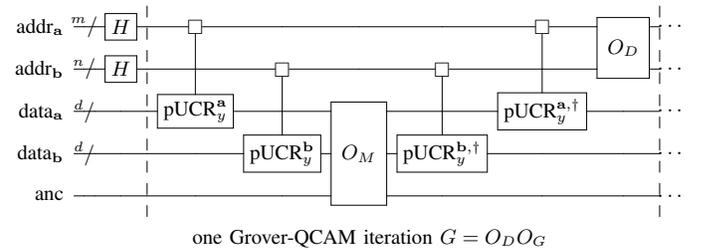

\resizebox{\columnwidth}{!}{%
\(\small
\quad \qquad 
\begin{myqcircuit}
\lstick{\text{addr}_{\mathbf{a}}}  & ^{m}{/} \qw & \qw & \gate{H} \barrier[0em]{4}&\qw  & \ctrlsq{2}                  & \qw                         & \qw                & \qw & \ctrlsq{2}& \multigate{1}{O_D} \barrier[0em]{4}& \qw &  \cdots \\
\lstick{\text{addr}_{\mathbf{b}}}  & ^{n}{/} \qw & \qw & \gate{H} &\qw& \qw                         & \ctrlsq{2}                  & \qw                & \ctrlsq{2} & \qw & \ghost{O_D} & \qw & \cdots\\
\lstick{\text{data}_{\mathbf{a}}}  & ^{d}{/} \qw & \qw & \qw   &\qw   & \gate{\pUCRy^\mathbf{a}}             & \qw                         & \multigate{2}{O_M} & \qw & \gate{\pUCRy^{\mathbf{a},\dagger}} & \qw & \qw & \cdots\\
\lstick{\text{data}_{\mathbf{b}}}  & ^{d}{/} \qw & \qw & \qw   &\qw   & \qw                         & \gate{\pUCRy^\mathbf{b}}             & \ghost{O_M}        & \gate{\pUCRy^{\mathbf{b},\dagger}} & \qw & \qw & \qw & \cdots\\
\lstick{\text{anc}}     & \qw & \qw      & \qw & \qw   &\qw                      & \qw                         & \ghost{O_M}        & \qw & \qw & \qw & \qw & \cdots\\
 & & & &&&&~~~\text{one Grover-QCAM iteration $G = O_D O_G$}   \\
\end{myqcircuit}
\)
}
\vspace{1mm}
\caption{Circuit diagram illustrating the initial superposition over the address spaces and the first Grover iteration $G$ for QCAM. See text for further details.
}
\label{fig:grover-qcam}
\end{figure}

The circuit diagram for the first Grover-QCAM iteration in the QCAM circuit is shown in Fig.~\ref{fig:grover-qcam}. It consists of the following elements:

\begin{enumerate}
    \item[(1)] Hadamard gates on the $n + m$ qubits representing the data registers for $\mathbf{a}$ and $\mathbf{b}$.
    \item[(2)] A pair of pUCR gates, effectively loading the data; at this stage the  4 multi-qubit registers in Fig.~\ref{fig:grover-qcam} are in the state
    \begin{equation}
        \frac{1}{\sqrt{NM}} \sum_{i,j} \ket{i}\ket{j} \otimes \ket{a_i} \ket{b_j}.
    \end{equation}
    \item[(3)] A \emph{matching oracle} $O_M$ acting only on the data registers and the single ancilla as
    \begin{equation}
        O_M \ket{a_i}\ket{b_j}\ket{x} \mapsto (-1)^{\overline{a_i \oplus b_j}}\,\ket{a_i}\ket{b_j}\ket{x},
        \label{eq:OM}
    \end{equation}
    where $\oplus$ is a bitwise XOR and the overline indicates a NOT operation. The $O_M$ oracle adds a relative phase to all the states for which the data match.
    \item[(4)] The inverse of the pair of pUCR gates, effectively uncomputing the data encodings.
    \item[(5)] The Grover diffuser $O_D$ applied only on the address registers.
\end{enumerate}

The \emph{Grover oracle} $O_G$ consists of all circuit elements described in steps (2)-(4). The \emph{Grover iterator}, $G = O_D O_G$,
is the product of steps (2)-(5).

\begin{figure}[htbp]
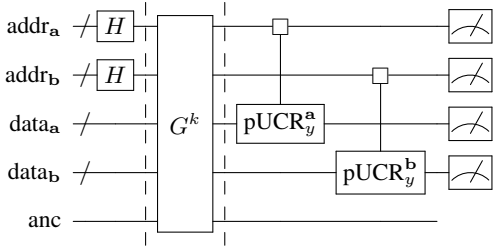

\centering
\resizebox{0.75\columnwidth}{!}{%
\(\small
\quad \qquad 
\begin{myqcircuit}
\lstick{\text{addr}_{\mathbf{a}}}  & {/} \qw & \gate{H} \barrier[0em]{4}& \qw  & \multigate{4}{G^k}\barrier[0em]{4} & \qw & \ctrlsq{2} & \qw & \qw & \meter\\
\lstick{\text{addr}_{\mathbf{b}}}  & {/} \qw & \gate{H}                 & \qw  & \ghost{G^k}        & \qw & \qw &\ctrlsq{2} & \qw & \meter\\
\lstick{\text{data}_{\mathbf{a}}}  & {/} \qw & \qw                      & \qw  & \ghost{G^k}        & \qw & \gate{\pUCRy^\mathbf{a}} & \qw & \qw & \meter\\
\lstick{\text{data}_{\mathbf{b}}}  & {/} \qw & \qw                      & \qw  & \ghost{G^k}        & \qw & \qw & \gate{\pUCRy^\mathbf{b}} & \qw & \meter\\
\lstick{\text{anc}}                & \qw & \qw                      & \qw  & \ghost{G^k}        & \qw & \qw & \qw & \qw\\
\end{myqcircuit}
\)
}
\vspace{1mm}
\caption{Circuit diagram illustrating the initial equal superposition over the address spaces, $k$ Grover iterations $G^k$ for QCAM to amplify the addresses of the matches in the sequences, and the final data loading circuits with measurements.}
\label{fig:grover-qcam-data}
\end{figure}

The complete Grover-based QCAM algorithm 
is made from repeating the Grover iterator $k$ times, $G^k H \ket{0}$, where $k$ depends the number of possible solutions ($NM$, i.e., number of possible pairs of addresses)
and the true number of matches (solutions) $K$. The latter quantity is unknown a priory and is input dependent. In Sec.~\ref{sec:HEQC}
we describe a hardware-efficient procedure to estimate $K$ on a QPU with a dedicated circuit. 
At that point, we'll also discuss how to
compute the ground truth number of iterations $k$ from the number of solutions  $K$.
If the Grover iterator is repeated $k$ times, there will be an $O(1)$ probability that the registers labeled $\text{addr}_{\mathbf{a}}$ and $\text{addr}_{\mathbf{b}}$ contain only pairs of addresses of  matches in sequences $\mathbf{a}$ and $\mathbf{b}$. As the Grover iterator treats the registers $\text{data}_{\mathbf{a}}$, $\text{data}_{\mathbf{b}}$ and anc as ancillary workspaces which are uncomputed in every iteration, the data loading pUCR circuit(s) have to run an additional time, as shown in Fig.~\ref{fig:grover-qcam-data}, in order to also measure the data values for corresponding matched addresses.
We note that strictly speaking only one pUCR circuit is required to encode only one of the two sequences, but the second one will be executed concurrently and can be used to verify that a match has indeed been observed.
Furthermore, this is a constant overhead which does not increase the overall algorithm complexity of QCAM.

\begin{figure}[htbp]
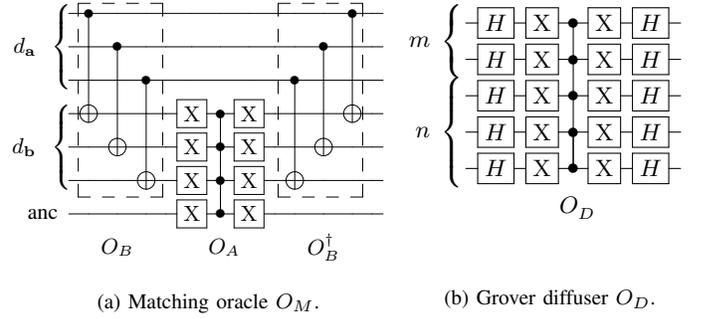

\subfloat[][Matching oracle $O_M$.]{
\resizebox{0.55\columnwidth}{!}{%
\(\small
\quad 
\begin{myqcircuit}
&&                               & \ctrl{3} & \qw & \qw & \qw & \qw & \qw & \qw & \qw & \qw & \qw & \ctrl{3} &  \qw  &\qw\\
\lstick{d_\mathbf{a}} && & \qw & \ctrl{3} & \qw & \qw & \qw & \qw & \qw & \qw & \qw & \ctrl{3} & \qw & \qw & \qw\\
&&                               & \qw & \qw & \ctrl{3} & \qw & \qw & \qw & \qw & \qw & \ctrl{3} & \qw & \qw & \qw & \qw\\
 && & \targ & \qw & \qw & \qw & \gate{\mathrm{X}} & \ctrl{1} & \gate{\mathrm{X}} & \qw & \qw & \qw & \targ & \qw& \qw\\
 \lstick{d_\mathbf{b}} && & \qw & \targ & \qw & \qw & \gate{\mathrm{X}} & \ctrl{1} & \gate{\mathrm{X}} & \qw & \qw & \targ & \qw & \qw & \qw\\
 && & \qw & \qw & \targ & \qw & \gate{\mathrm{X}} & \ctrl{1} & \gate{\mathrm{X}} & \qw & \targ & \qw & \qw & \qw & \qw\\
 && \lstick{\text{anc}} & \qw & \qw & \qw & \qw & \gate{\mathrm{X}} & \ctrl{-1} & \gate{\mathrm{X}} & \qw & \qw & \qw & \qw & \qw & \qw
{\gategroup{1}{3}{3}{5}{0.7em}{\{}}
{\gategroup{4}{3}{6}{5}{0.7em}{\{}}
{\gategroup{1}{4}{6}{6}{0.7em}{--}}
{\gategroup{1}{12}{6}{14}{0.7em}{--}}\\
&&&&O_B  &&& &~O_A  &&&&O_B^{\dagger} \\ 
\\
\end{myqcircuit}
\)
}
}
\subfloat[][Grover diffuser $O_D$.]{
\resizebox{0.42\columnwidth}{!}{%
\(\small
\ \quad
\begin{myqcircuit}
    && & \gate{H} & \gate{\mathrm{X}}  & \ctrl{1} & \gate{\mathrm{X}} & \gate{H} & \qw \\
\lstick{\raisebox{1.5em}{$m$}} &&    & \gate{H} & \gate{\mathrm{X}} & \ctrl{1} & \gate{\mathrm{X}} & \gate{H}  & \qw\\
    && & \gate{H} & \gate{\mathrm{X}} & \ctrl{1} & \gate{\mathrm{X}} & \gate{H}  & \qw\\
\lstick{n} &&    & \gate{H} & \gate{\mathrm{X}} & \ctrl{1} & \gate{\mathrm{X}} & \gate{H}  & \qw\\
    && & \gate{H} & \gate{\mathrm{X}} & \ctrl{-1} & \gate{\mathrm{X}}  & \gate{H}  & \qw \\
    &&&&&~O_D^{\phantom{\dagger}} \\
    &&\\
{\gategroup{1}{2}{2}{4}{0.2em}{\{}}
{\gategroup{3}{2}{5}{4}{0.2em}{\{}}
\end{myqcircuit}
\)
\vspace{1000em}
}
}
\caption{The circuits implementing the matching oracle $O_M$  and the Grover diffuser $O_D$ .
}
\label{fig:oracles}
 \end{figure}

For completeness, we  describe in Fig.~\ref{fig:oracles} how to implement the matching oracle $O_M$ (Eq.~\eqref{eq:OM}) and the
Grover diffuser $O_D$ (Eq.~\eqref{eq:OD}) as quantum circuits.  We note that the matching oracle assumes that the bit-depths
satisfy $d_\mathbf{a} = d_\mathbf{b}$, with the subscripts included for clarity.
The Grover diffuser does not impose that $m = n$, i.e., it will handle correctly sequences
of different lengths.


The matching oracle~\eqref{eq:OM}, shown in Fig.~\ref{fig:oracles}(a),
acts in three steps.
In the first step, $O_B$, acts on the two data registers
$\ket{i}$ and $\ket{j}$ of equal size $d$ as
\begin{equation}
    O_B \ket{i}\ket{j}\ket{0} \mapsto \ket{i} \ket{\overline{i \oplus j}}\ket{0}.
    \label{eq:OB}
\end{equation}
In words, $O_B$ sets all qubits in the second qubit register to 1 only if the bit strings in $i$ and $j$ match, the  ancilla qubit remains unchanged.
A second oracle, $O_A$, adds a relative phase conditioned on the second register being 1 for all qubits,
\begin{equation}
    O_A \ket{i}\ket{j}\ket{0} \mapsto (-1)^{\overline{j}} \ket{i} \ket{j}\ket{0},
    \label{eq:OA}
\end{equation}
where $\bar{j}$ indicates a bit-wise negation.

This step requires the ancilla qubit as workspace.
Finally, we restore the states of both input registers $\ket{i}\ket{j}$ by uncomputing with $O_B^{\dagger}$. The full unitary for the matching oracle becomes $O_M = O_B^{\dagger} O_A O_B$, which implements Eq.~\eqref{eq:OM}.

It is well-known that the Grover diffuser (Fig.~\ref{fig:oracles}(b)) implements the
reflection operator,
\begin{equation}
O_D = H^{\otimes(n + m)} (2 \ket{0} \bra{0} - I) H^{\otimes(n + m)},
\label{eq:OD}
\end{equation}
which reflects over the state of equal superposition, $\ket{+} \equiv H^{\otimes(n + m)}\ket{0}$.

In general, the Grover-based QCAM scheme presented in Fig.~\ref{fig:grover-qcam} can be bootstrapped to search for matches in more than two sequences. In Sec.~\ref{sec:dna}, we apply the QCAM primitive to match k-mers in two DNA sequences on a simulated QPU.

\section{HEQC: Hardware Efficient Quantum Counting Algorithm}\label{sec:HEQC}

In this section, we address one of the key questions relevant to every Grover problem: How to efficiently estimate the near-optimal number of iterations $k$?
As we will show later in this section (Eq.~\eqref{eq:iter}), $k$ can be determined from the phase of the Grover iterator, $\theta$,  acting in the Grover subspace on a specific, easy to prepare state. 
At the same time, knowing the phase of the Grover
iterator also allows one to compute the number of solutions to the Grover search problem if the number of possible solutions (size of search space) is known.
The standard approach for computing the phase of the Grover iterator $G$ is known as the 
\emph{quantum counting algorithm}~\cite{qcounting} because it also counts the number of solutions. The known quantum counting algorithm requires full quantum phase estimation~\cite{qpe} on the Grover iterator $G$ to determine its phase.

In this section, we introduce a hardware efficient quantum counting (HEQC) algorithm that --under minor assumptions-- only requires the estimation of a single squared expectation value of one Grover oracle $O_G$ acting on a specific initial state. HEQC is applicable to any Grover problem, including QCAM.
For ease of notation and adhering to existing conventions~\cite{NCbook}, we overload and redefine some variables used in  the previous section.
We emphasize that a number of simplified quantum counting algorithms~\cite{wie2019simpler,aaronson2020quantum} (compared to the standard method~\cite{qcounting} which requires phase estimation) have been introduced in the last few years and may prove more efficient than HEQC in certain regimes.

Let $G$ indicate a Grover problem and iterator over a search set $S \cong \left[N\right]$ of cardinality $N$, and let $B \subseteq S$ be the set of solutions, with $|B| = M$ the number of solutions to the Grover problem. Let's assume the problem is non-trivial, i.e.: $B \ne S$ and $| B \cap S |>0 $.

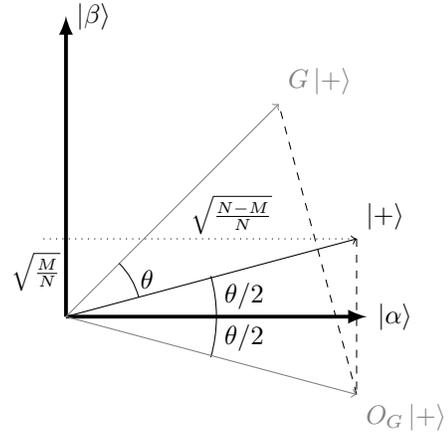
\begin{figure}[htbp!]
\centering
\begin{tikzpicture}[scale=4]
    \coordinate  (origin) at  (0,0);
    \draw[line width=0.5mm,-latex] (origin) -- (1,0) node[below,right] {$\ket{\alpha}$} ;
    \draw[line width=0.5mm,-latex] (origin) -- (0,1) node[above,right] {$\ket{\beta}$} node[left=0.5cm] {};
    \draw[->] (origin) -- (0.9659,0.2588) node[above right] {$\ket{+}$};
    \draw (0.5,0) arc (0:15.7:0.5) node[midway,right] {$\theta/2$};
    \draw (0.5,0) arc (0:-15.7:0.5) node[midway,right] {$\theta/2$};
    \draw[dotted] (-0.075, 0.2588) -- (0.9659,0.2588);
    \node[] at (-0.1,0.15) {{\footnotesize$\sqrt{\frac{M}{N}}$}};
    \node[] at (0.55, 0.34){{\footnotesize $\sqrt{\frac{N-M}{N}}$}};
    \draw[dashed] (0.9659,0.2588) -- (0.9659,-0.2588);
    \draw[dashed] (0.9659,-0.2588) -- (0.70711,0.70711);
    \draw[gray, ->] (origin) -- (0.70711,0.70711) node[above right] {$G\ket{+}$};
    \draw (0.2415,0.065) arc (15:45:0.25) node[midway, right] {$\theta$};
    \draw[gray, ->] (origin) -- (0.9659,-0.2588) node[below right] {$O_G \ket{+}$};
\end{tikzpicture}
\caption{Geometric representation of the Grover subspace $\mathcal{G}$, the starting state $\ket{+}$, the action of the Grover oracle $O_G\ket{+}$ and Grover operator $G\ket{+}$.\label{fig:grover_geom}
}
\end{figure}

The states $\ket{\alpha}$ and $\ket{\beta}$, defined by,
\begin{equation}
\begin{split}
\ket{\alpha} &= \frac{1}{\sqrt{N-M}} \sum_{x \notin B} \ket{x},\\
\ket{\beta} &= \frac{1}{\sqrt{M}} \sum_{x \in B} \ket{x},
\end{split}
\end{equation}
are equal superpositions over sets $S \setminus B$ and $B$, respectively.
These vectors form an orthonormal basis of the two-dimensional \emph{Grover subspace} $\mathcal{G} = \text{span}\{ \ket{\alpha}, \ket{\beta} \}$.
We call $\mathcal{G}$ the Grover subspace as it is well-known that the
Grover iterator $G$ acts as a planar rotation on $\mathcal{G}$~\cite{grover,NCbook}, i.e.,
\begin{equation}
    G\vert_{\mathcal{G}} = \begin{bmatrix} \cos \theta & - \sin \theta \\ \sin \theta & \phantom{-}\cos \theta \end{bmatrix},
\end{equation}
with the phase $\theta$ directly related to $M$ and $N$.
The equal superposition over $S$,
\begin{equation}
\begin{split}
\ket{+} & = \frac{1}{\sqrt{N}} \sum_{x \in S} \ket{x}, \\
         & = \sqrt{\frac{N-M}{M}} \ket{\alpha} + \sqrt{\frac{M}{N}} \ket{\beta}, \\
         & = \cos \frac{\theta}{2} \ket{\alpha} + \sin  \frac{\theta}{2} \ket{\beta},
\end{split}
\end{equation}
lies in the Grover subspace, $\ket{+} \in \mathcal{G}$.
The geometric picture of $\mathcal{G}$ is shown in Fig.~\ref{fig:grover_geom}
and highlights that $G\vert_{\mathcal{G}}$ is constructed as a product of two reflections, $O_D O_G$, along respectively the $\ket{+}$ and $ \ket{\alpha}$ states. It follows that the phase $\theta$ of $G\vert_{\mathcal{G}}$ is double the
angle between $\ket{\alpha}$ and $\ket{+}$.
Hence, if we have computed $\theta$, we can infer
\begin{equation}
M = N \sin^2 \left(  \frac{\theta}{2}\right),
\label{eq:estM}
\end{equation}
as $N$ is assumed to be known.

\begin{figure}[htbp!]
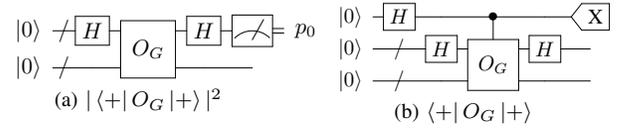

\centering
\subfloat[][$|\bra{+}O_G\ket{+}|^2$]{
\resizebox{0.46\columnwidth}{!}{%
\(
\qquad \quad
\begin{myqcircuit}
\lstick{\ket{0}} & {/} \qw & \gate{H} & \multigate{1}{O_G} & \gate{H} & \meter & \rstick{p_0} \cw\\
\lstick{\ket{0}} & {/} \qw  & \qw      & \ghost{O_G} & \qw & \qw\\
\end{myqcircuit}
\)
}
}
\subfloat[][$\bra{+}O_G\ket{+}$]{
\resizebox{0.46\columnwidth}{!}{%
\(
\qquad \quad
\begin{myqcircuit}
\lstick{\ket{0}} & \gate{H} & \qw & \ctrl{1} & \qw & \measuretab{\mbox{X}} \\
\lstick{\ket{0}} & {/} \qw & \gate{H} & \multigate{1}{O_G} & \gate{H} & \qw\\
\lstick{\ket{0}} & {/} \qw  & \qw      & \ghost{O_G} & \qw & \qw \\
\end{myqcircuit}
\)
}
}
\caption{(a) HEQC circuit for determining the phase of the Grover oracle $O_G$ (and consequently Grover iterator $G$) in case $M \leq N/2$: the first 
multi-qubit register is the Grover search register, the second register is an (optional) ancillary work register. 
The phase $\theta$ can be directly estimated from measuring $|\bra{+}O_G\ket{+}|^2$, which corresponds to the probability $p_0$ of measuring the zero string on the search register. In case $M \leq N$, a Hadamard test circuit (b) is required to determine the sign of the overlap. Measuring the real part of the overlap suffices.\label{fig:HEQC}
}
\end{figure}

The key observation for HEQC is that we can directly estimate $\theta$
from,
\begin{equation}
\theta = \arccos \bra{+} O_G \ket{+},
\label{eq:theta}
\end{equation}
which avoids full phase estimation and also does not require the implementation of the Grover diffuser $O_D$.
Furthermore, if we know $M \leq N/2$, then $\theta \in [0, \pi/2]$ and
$\bra{+}O_G\ket{+} \geq 0$. This means we can compute $\theta$ from the squared
overlap $|\bra{+}O_G\ket{+}|^2$ which can be estimated on a QPU without introducing ancillary qubits using a circuit shown in Fig.~\ref{fig:HEQC}(a).
If all we know is that $M \leq N$, then $\theta \in [0, \pi]$ and the overlap can be negative. In this case, we have to measure the overlap to retrieve the sign and can
do so using a Hadamard test circuit in Fig.~\ref{fig:HEQC}(b).

Once we have computed $\theta$ by applying Eq.~\eqref{eq:theta} to the measurement result of the circuit shown in Fig.~\ref{fig:HEQC}, HEQC proceeds in the same way as regular quantum counting to estimate the number of iterations $k$.
This means that, starting from $\ket{+}$, we need to rotate over an angle of $(\pi - \theta)/2$ to arrive at $\ket{\beta}$ (Fig.~\ref{fig:grover_geom}). As $G$ takes discrete steps over angles $\theta$, the estimate for the number of Grover iterations thus
becomes:
\begin{equation}
    k = \texttt{ROUND}\left( \frac{\pi - \theta}{2 \theta} \right),
    \label{eq:iter}
\end{equation}
where $\texttt{ROUND}(x)$ denotes rounding of $x$ to the \emph{closest integer}.
Note that, Eq.~\eqref{eq:iter} does only guarantee that we stop the Grover iteration at a state which, when measured, yields a solution to the Grover problem with probability at least 0.5.

\begin{figure}[htbp]
\centerline{\includegraphics[width=.9\linewidth]{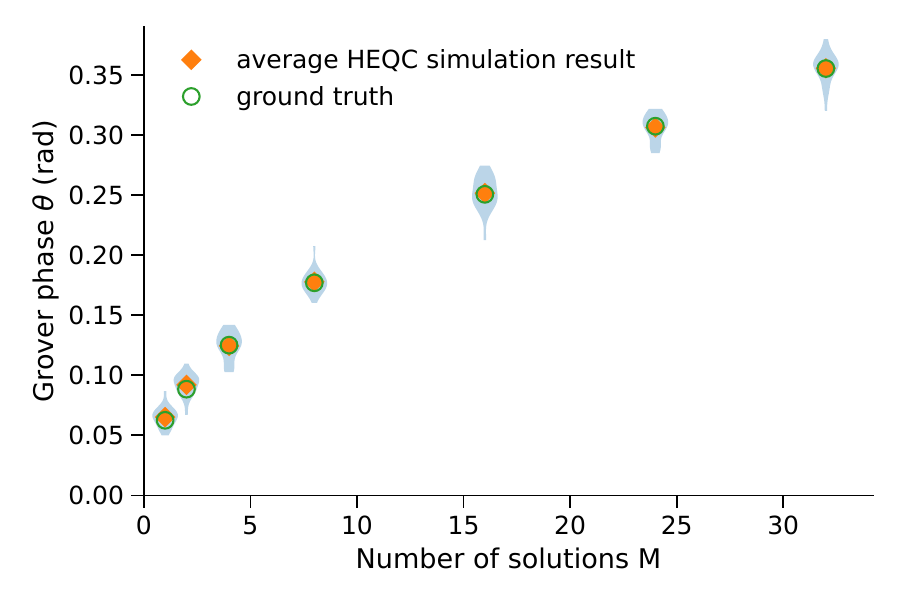}}
   \caption{Simulation results of the HEQC algorithm to compute the Grover phase $\theta$. $N$ is kept fixed at 1024 and $M$ is varied between 1 and 32. Simulations for each value of $M$ was repeated 21 times, using 2000 shots each time. The blue violin plot show the spread of reconstructed $\theta$ values with the orange diamond marking the mean. The green circle marks the ground truth computed with Eq.~\eqref{eq:estM}. }
\label{fig:reco-theta}
 \end{figure}

We have just shown that HEQC can estimate the phase of the Grover iterator, the number of solutions to a Grover search problem (Eq.~\eqref{eq:estM}) and the near-optimal number of iterations (Eq.~\eqref{eq:iter}) while only having to measure a single observable of $O_G$. Next, we demonstrate the performance of the proposed HEQC algorithm for QCAM using the Qiskit shot-based circuit simulator. We generate two random sequences of 32 8-bit integers and use the Grover-QCAM oracle, $O_G$, shown in Fig.~\ref{fig:grover-qcam}, to encode and match them.
The complete HEQC circuit requires 27 qubits.
The size of  search space is $N=1024$ and we assumed that exactly $M$ pairs of matched values existed in the 2 input sequences. We varied $M$ from 1 to 32 and for each value of $M$, we simulated 21 circuits for different random sequences using 2000 shots each. The $\theta$ was computed using Eq.~\eqref{eq:theta} based on the estimated overlap.
On average, we observe a good agreement with the ground truth in Fig.~\ref{fig:reco-theta}, with a modest spread on the random realizations.


\section{Jaccard similarity for DNA}
\label{sec:dna}

Deoxyribonucleic acid (DNA) is the primary molecule used to store genetic information in living organisms. DNA consists of polynucleotide chains built from the four nucleotides: adenine, cytosine, thymine, and guanine, abbreviated A, C, T, and G, respectively.
 Recent advances in DNA sequencing technology, in particular 
  metagenomic sequencing, generate thousands to tens of millions of unlabelled DNA sequences. To make sense of all of these sequences, scientists compare the unlabelled DNA sequences to DNA sequences for which the species is known. A common way to do this is to first break down the individual sequences into a set of \textit{k}-mers, i.e. overlapping substrings of length \textit{k}, see Table~\ref{tab:kmer_example}. These sets are then compared using the Jaccard similarity metric \cite{Jaccard}. This method is computationally expensive and does not trivially map to modern computer architectures.

\begin{table}[htbp]
\begin{center}
  \caption{Example DNA sequence of length 8 with all unique $k$-mers (in order of appearance) for $k=1,2,3$. The maximum number of unique $k$-mers equals $4^k$. Only all 1-mers appear in the example sequence.}
\resizebox{\columnwidth}{!}{%
\begin{tabular}{rlc}
\hline 
DNA sequence & TGTCGAAA & max\\
\hline 
1-mers& T, G, C, A & $4^{1} $\\
2-mers & TG, GT, TC, CG, AA & $4^2$\\
3-mers & TGT, GTC, TCG, CGA, GAA, AAA& $4^3$\\
\hline
\end{tabular}
}
\label{tab:kmer_example}
\end{center}
\end{table}
 
The {\it Jaccard similarity} metric between two sets $A$, $B$ is defined as the size of their intersection divided by the size of their union,
\begin{equation}
\label{eq:jacc-def}
J(A,B) = \frac{ |A \cap B|}{|A \cup B|} = \frac{ |A \cap B|}{|A| + |B| - |A \cap B|}.
\end{equation}
The complexity of this calculation, $O(|A| + |B|)$, is determined by the set intersection operation. Executing this linear operation on modern computer architectures is frequently limited due to the size of large DNA datasets, which often do not fit into the memory available on typical compute nodes.
We construct a QCAM quantum circuit that samples from the intersection $A \cap B$ \emph{with duplication}.
Computing the cardinality $|A \cap B|$ is achieved through classical post-processing of the measurement results, which includes rejecting duplicates. 
The cardinalities $|A|$ and $|B|$ are assumed to be known or easily computable classically, such that $J(A,B)$ can be directly computed from Eq.~\eqref{eq:jacc-def}.
In contrast, the QCAM matching algorithm only requires a number of qubits that scales logarithmicly with the length of the DNA sequences. Instead, the bottleneck for QCAM lies in the \emph{depth} of the data loading circuits.

\subsection{DNA encoding in bit-strings}

Two bits suffice to encode the  4 types of nucleotides ($A$, $T$, $G$, $C$). For a $k$-mer of length $k$ the $2k$ bits are needed.
On a quantum computer, we can encode a nucleotide in the computational basis states.
Encoding the 4 different nucleotides requires 2 qubits, 
\begin{equation}
    \ket{A} := \ket{00}, \ket{T} := \ket{01}, \ket{G} := \ket{10}, \ket{C} := \ket{11},
\end{equation}
which is analogous to the classical encoding.
Similarly, we encode a $k$-mer in the computational basis of $2k$ qubits, e.g., $4$-mers require 8 qubits,
\begin{equation}
    \ket{TACT} := \ket{01001101}, \ket{GATG} := \ket{10000110}, \ldots
\end{equation}

To store one DNA strand in an NEQR state (Eq.~\eqref{eq:neqr}) as a sequence of $2^n$ overlapping $k$-mers (with duplications),
we need $n$ address qubits and $2k$ data qubits for a total of $n+2k$.
For example, the DNA strand $ATGATGA$ of length 7 can be represented in an NEQR state as a sequence of 4 4-mers:
\begin{equation}
    \frac{1}{2}(\ket0\ket{ATGA} + \ket1\ket{TGAT} + \ket2\ket{GATG} + \ket3\ket{ATGA}).
    \label{eq:kmerenc}
\end{equation}
The encoding of this DNA strand requires $n=2$ address qubits and $d=8$ data qubits.
We note that \eqref{eq:kmerenc} does include duplicate 4-mers at addresses $\ket0$ and $\ket3$.

It follows that our Grover-based QCAM algorithm, introduced in Sec.~\ref{sec:qcam}, requires $n + m + 4k + 1$ qubits to search for matches
in two sequences of $k$-mers $\mathbf{a}$ and $\mathbf{b}$  of length $2^n$ and $2^m$, respectively.

\subsection{Results}

We simulate the Jaccard index computation through QCAM for various lengths of DNA strands and $k$-mers.
For simplicity, we always use two DNA strands of the same length.
The first strand is randomly generated and the second one is a copy of the first one with 10\% random mutations added.  
The number of Grover iterations can be determined through the HEQC algorithm described in Sec.~\ref{sec:HEQC}.
 
Table~\ref{tab:dna} shows examples of simulated results for  several choices of DNA sample lengths and  $k$-mers sizes.  
The largest simulated circuit has 33 qubits.
 For all simulated cases the QCAM result is exact for the chosen number of shots.
 
\begin{table}[htbp]
\caption{Simulated computation of Jaccard index for random DNA samples of different lengths. }
\begin{center}
\resizebox{\columnwidth}{!}{%
\begin{tabular}{P{8mm}P{7mm}P{6mm}P{8mm}P{10mm}P{7mm}P{8mm}}
\hline
\textbf{$k$-mer length} & \textbf{DNA length} & total qubits & addr. pairs$\dag$& unique $k$-mers$\dag$ & Grover iter$\dag$ & shots \\
\hline
3 & 128 & 27 &  350 & 60 & 5& 11,000  \\
4 & 128 & 31 & 130 & 80 & 8 & 4,000 \\
5 & 64 & 33 & 38 & 34 & 8 & 1,100  \\
6 & 16 & 33 & 11 & 8 & 3 & 330  \\
\hline
\end{tabular}
}
\label{tab:dna}
\end{center}
$\dag$: typical result
\end{table}

Fig.~\ref{fig:dna64} shows an example input and output of this Jaccard index algorithm
 for the choice of DNA length of 64 bases with  the 4-mer tiling.  
Two input DNA strands are labeled as  A and B.
The {\it common} sequence flags the differences between this 2 strands using asterisks. 
A subset of 4-mers belonging to $A \cap B $ is listed below it. For this particular input  the quantum algorithm found  56, 55, and 36  distinct 4-mers in samples, A, B, and $A \cap B $  , respectively. Hence the computed Jaccard index is of 0.48.

\begin{figure*}[htbp]
\begin{lstlisting}
sample A : CAATGAATGTGTCCACTGGATTGACAGTCTGGGATGAGCGCACTTCACGGATTGTTCTTGCCGAACCC
sample B : CAATGAATGTTTGCACAAGATTGACAGCCGGGGATTAGCGCACTTCACGGATTGCTCTTGCCGAACCC
common   : CAATGAATGT*T*CAC**GATTGACAG*C*GGGAT*AGCGCACTTCACGGATTG*TCTTGCCGAACCC

4-mers in $A \cap B $  : {'AACC', 'AATG', 'ACAG', 'ACGG', 'ACTT', 'AGCG', ... }
4-mers counts  $|A|$=56,  $|B|$=55,  $|A \cap B| $=36,   Jaccard Index=0.480
\end{lstlisting}
\caption{Pair of DNA samples of length 64 with Jaccard index of 0.48 assuming overlapping 4-mers tailing. Asterisks in the `common' strand mark mutations differentiating strand A from B.}
\label{fig:dna64}

 \end{figure*}

\section{Discussion}
\label{sec:disc}

In this work, we use the pUCR gates first introduced for the \qbart\ circuits for NEQR data encodings~\cite{qbart}, 
to construct a highly-optimized Grover search oracle applicable for finding matches in two data sequences of bit-strings.
In analogy to classical content-addressable memory~\cite{cam}, which also allows finding matches between data values loaded into memory,
we call this application QCAM.
In contrast to classical CAM, which returns all the matches in a predictable order, QCAM randomly retrieves only one of possibly many matches every single shot. However, QCAM does deliver the typical Grover
quadratic speedup over unstructured search, without requiring special circuitry as in the case of classical CAM.

The second contribution of our work is the introduction of HEQC, the hardware efficient implementation of the quantum counting algorithm~\cite{qcounting}.  HEQC allows one to retrieve the number of solutions to the Grover search problem from the measurement of the single observable $\bra{+}O_G\ket{+}$ using a relatively shallow circuit \emph{and} consequently, to compute a near-optimal number of Grover iterations, given some specific search problem.
Our simulation results of the HEQC circuit show a good agreement with actual Grover phase angle.

Finally, we applied the QCAM algorithm to compute the Jaccard index between two strands of DNA and demonstrated an end-to-end proof of principle shot-based simulation for small-scale, random DNA strands.
In practice, Jaccard similiarity based analyses of DNA typically use larger $k$-mers than we have used here. For example, Sourmash \cite{sourmash} distributes precomputed databases generated using lengths \textit{k}=21, \textit{k}=31, or \textit{k}=51.
Furthermore, the  real metagenomic samples  can contain millions of unique   $k$-mers. Application of QCAM for such scenarios would require  QPUs with $O(200)$ qubits and  fidelity appropriate for the circuit depth  of $O(10^{12})$ of entangling operations. 

In future work, we will investigate the possibility of further optimizing the quantum algorithm to compute the Jaccard similarity. E.g., we could reduce two DNA sequences to two sets of unique $k$-mers in a classical pre-processing at a cost linear in the DNA length. Then, HEQC would give us the phase 
of the Grover iterator $\theta$, directly allowing the computation of $|A \cap B|$ without having to run all $k$ Grover iterations and without ever extracting the set of  matched k-mers $A \cap B$. 
The trade-off between the  reduced number of iterations and retrieving information from the expectation value rather than from the most probable bit-strings
remains to be studied.
Notably other problems of interest, such as graph traversing, require more detailed output about matched vertices which would be provided only by QCAM.

A second future extension to optimize the Jaccard index computation on quantum hardware includes the development of quantum circuits that generate a $k$-mer encoding of a DNA strand, as shown in Eq.~\eqref{eq:kmerenc}, directly from a 1-mer encoding.
We will investigate the use of \emph{cyclic shift} circuitry to achieve this goal.
This alternative approach has the potential to reduce the redundancy in the pUCR data loading circuits as all the information to generate the complete $k$-mer sequence will only have to be loaded once as a 1-mer sequence.


\section*{Acknowledgment}
This research used resources of the National Energy Research Scientific Computing Center (NERSC), a U.S. Department of Energy Office of Science User Facility located at Lawrence Berkeley National Laboratory, operated under Contract No. DE-AC02-05CH11231.


\end{document}

%% file: custom-style.tex
\usepackage{rotating}

\newcommand{\UCR}[1][]{
    \ifthenelse{ \equal {#1} {} }
        {\text{UCR}}
        {\text{UCR}^{(#1)}}
}

\newcommand{\UCRy}[1][]{\UCR[#1]_{y}}
\newcommand{\pUCRy}[1][]{\text{p}\UCRy[#1]}
\newcommand{\psifrqi}[1][]{
    \ifthenelse{ \equal {#1} {} }
    {\ket{\psi_{\text{FRQI}}}}
    {\ket{\psi_{\text{FRQI}}(#1)}}
}
\newcommand{\psiqcrank}[1][]{
    \ifthenelse{ \equal {#1} {} }
    {\ket{\psi_{\text{qcrank}}}}
    {\ket{\psi_{\text{qcrank}}(#1)}}
}

\usepackage[braket]{qcircuit}

\usepackage{environ}
\newcommand{\myqctmp}[2][0.25]{\Qcircuit @C=#2em @R=#1em @!R}
\NewEnviron{myqcircuit}[1][0.25]{\vcenter{\myqctmp[#1]{0.5} {\BODY}}}
\NewEnviron{myqcircuit*}[2]{\vcenter{\myqctmp[#1]{#2} {\BODY}}}

\newcommand{\controlsq}{*!<0em,.025em>-=-<0em>{\square}}
\newcommand{\ctrlsq}[1]{\controlsq \qwx[#1] \qw}

\usepackage{lipsum} 

\usepackage[switch]{lineno}

\usepackage{listings} 
\usepackage{algorithm}
\usepackage{algpseudocode}

\newcommand{\qbart}{\textbf{\texttt{QBArt}}}


%% file: main.bbl
\begin{thebibliography}{10}
\providecommand{\url}[1]{#1}
\csname url@samestyle\endcsname
\providecommand{\newblock}{\relax}
\providecommand{\bibinfo}[2]{#2}
\providecommand{\BIBentrySTDinterwordspacing}{\spaceskip=0pt\relax}
\providecommand{\BIBentryALTinterwordstretchfactor}{4}
\providecommand{\BIBentryALTinterwordspacing}{\spaceskip=\fontdimen2\font plus
\BIBentryALTinterwordstretchfactor\fontdimen3\font minus
  \fontdimen4\font\relax}
\providecommand{\BIBforeignlanguage}[2]{{%
\expandafter\ifx\csname l@#1\endcsname\relax
\typeout{** WARNING: IEEEtran.bst: No hyphenation pattern has been}%
\typeout{** loaded for the language `#1'. Using the pattern for}%
\typeout{** the default language instead.}%
\else
\language=\csname l@#1\endcsname
\fi
#2}}
\providecommand{\BIBdecl}{\relax}
\BIBdecl

\bibitem{cam}
K.~Pagiamtzis and A.~Sheikholeslami, ``Content-addressable memory (cam)
  circuits and architectures: a tutorial and survey,'' \emph{IEEE Journal of
  Solid-State Circuits}, vol.~41, no.~3, pp. 712--727, 2006.

\bibitem{grover}
\BIBentryALTinterwordspacing
L.~K. Grover, ``A fast quantum mechanical algorithm for database search,'' in
  \emph{Proceedings of the Twenty-Eighth Annual ACM Symposium on Theory of
  Computing}, ser. STOC '96.\hskip 1em plus 0.5em minus 0.4em\relax New York,
  NY, USA: Association for Computing Machinery, 1996, p. 212–219. [Online].
  Available: \url{https://doi.org/10.1145/237814.237866}
\BIBentrySTDinterwordspacing

\bibitem{qbart}
J.~Balewski, M.~G. Amankwah, R.~V. Beeumen, E.~W. Bethel, T.~Perciano, and
  D.~Camps, ``Quantum-parallel vectorized data encodings and computations on
  trapped-ions and transmon {QPU}s,'' 2023.

\bibitem{qcounting}
\BIBentryALTinterwordspacing
G.~Brassard, P.~H{\o}yer, and A.~Tapp, ``Quantum counting,'' in \emph{Automata,
  Languages and Programming}.\hskip 1em plus 0.5em minus 0.4em\relax Springer
  Berlin Heidelberg, 1998, pp. 820--831. [Online]. Available:
  \url{https://doi.org/10.1007/bfb0055105}
\BIBentrySTDinterwordspacing

\bibitem{Jaccard}
\BIBentryALTinterwordspacing
P.~Jaccard, ``The distribution of flora in the {A}lpine zone,'' \emph{New
  Phytologist}, vol.~11, no.~2, pp. 37--50, feb 1912. [Online]. Available:
  \url{https://doi.org/10.1111/j.1469-8137.1912.tb05611.x}
\BIBentrySTDinterwordspacing

\bibitem{QuRAM}
K.~Soni and A.~Malviya, ``Design and analysis of pattern matching algorithms
  based on quram processing,'' \emph{Arab J Sci Eng}, vol.~46, 2021.

\bibitem{Niroula}
P.~Niroula and Y.~Nam, ``A quantum algorithm for string matching,'' \emph{npj
  Quantum Inf}, vol.~7, 2021.

\bibitem{neqr}
Y.~Zhang, K.~Lu, Y.~Gao, and M.~Wang, ``{NEQR}: a novel enhanced quantum
  representation of digital images,'' \emph{Quantum Information Processing},
  vol.~12, 2013.

\bibitem{ucr}
M.~Möttönen, J.~J. Vartiainen, V.~Bergholm, and M.~M. Salomaa, ``Quantum
  circuits for general multiqubit gates,'' \emph{Physical Review Letters},
  vol.~93, 9 2004.

\bibitem{qpixl}
M.~G. Amankwah, D.~Camps, E.~W. Bethel, R.~V. Beeumen, and T.~Perciano,
  ``Quantum pixel representations and compression for {N}-dimensional images,''
  \emph{Scientific Reports}, vol.~12, p. 7712, 5 2022.

\bibitem{qpe}
A.~Y. Kitaev, ``Quantum measurements and the abelian stabilizer problem,''
  1995.

\bibitem{NCbook}
M.~A. Nielsen and I.~L. Chuang, \emph{Quantum Computation and Quantum
  Information: 10th Anniversary Edition}, 10th~ed.\hskip 1em plus 0.5em minus
  0.4em\relax USA: Cambridge University Press, 2011.

\bibitem{wie2019simpler}
C.-R. Wie, ``Simpler quantum counting,'' \emph{arXiv preprint
  arXiv:1907.08119}, 2019.

\bibitem{aaronson2020quantum}
S.~Aaronson and P.~Rall, ``Quantum approximate counting, simplified,'' in
  \emph{Symposium on Simplicity in Algorithms}.\hskip 1em plus 0.5em minus
  0.4em\relax SIAM, 2020, pp. 24--32.

\bibitem{sourmash}
\BIBentryALTinterwordspacing
C.~T. Brown and L.~Irber, ``sourmash: a library for {MinHash} sketching of
  {DNA},'' \emph{The Journal of Open Source Software}, vol.~1, no.~5, p.~27,
  sep 2016. [Online]. Available: \url{https://doi.org/10.21105/joss.00027}
\BIBentrySTDinterwordspacing

\end{thebibliography}
